\documentclass[showpacs,aps,amsmath,amssymb,twocolumn]{revtex4}

\usepackage{graphicx}
\usepackage{dcolumn}
\usepackage{bm}

\begin{document}

\title{The bound state Aharonov-Bohm effect around a cosmic string revisited}

\author{C. Filgueiras}
\affiliation{
Departamento de F\'{\i}sica, Universidade Federal de Pernambuco,\\
50670-901 Recife, PE, Brazil\\}
\author{Fernando Moraes}
\affiliation{Departamento de F\'{\i}sica, CCEN,  Universidade Federal 
da Para\'{\i}ba, Cidade Universit\'{a}ria, 58051-970 Jo\~ao Pessoa, PB,
Brazil}

\begin{abstract}
In this article we observe that the self-adjoint extension of the Hamiltonian of a  particle moving around a shielded cosmic string gives rise to a gravitational analogue of the bound state Aharonov-Bohm effect.

\end{abstract}


\maketitle
When a quantum particle is confined between two impenetrable concentric cylindrical walls, threaded by a magnetic flux tube along their common axis, its energy spectrum presents a dependence on the flux. This is the so-called bound state Aharonov-Bohm effect \cite{peshkin}. A gravitational analogue of this effect, with a cosmic string replacing the flux tube, has been studied in references \cite{valdir,kaluza}. This is a topological effect which also appears around disclinations in elastic solids \cite{sergio}. In this article we revisit the bound state gravitational Aharonov-Bohm effect from the point of view of self-adjoint extensions of the Hamiltonian, without need of a confining wall. 

Fermions in the presence of an Aharonov-Bohm field may have a bound state appearing from the self-adjoint extension of the Dirac Hamiltonian, as shown in \cite{gerbert}, for a specific range of the extension parameter. As shown in the detailed analysis of Kay and Studer \cite{kay}, a bound state may appear for a quantum particle moving about a cosmic string due to the self-adjoint extension of its Hamiltonian. On the other hand, the presence of walls may also lead to interesting effects related to the self-adjoint extensions as well \cite{french}. In this article we analyze the idealized situation of a free non-relativistic quantum particle in the presence of a cosmic string and therefore, moving in the background spacetime of metric
\begin{equation}
ds^2=dt^2-dz^2-d\rho^2-\alpha^2\rho^2 d\theta^2, \label{metric}
\end{equation}
shielded by an impenetrable cylindrical wall, of radius $a$, concentric with the string. In equation (\ref{metric}) the angle $\theta$ may vary in the range $[0,2\pi]$ and the parameter $\alpha=1-4G\mu<1$, where $\mu$ is the string linear mass density,  characterizes the string. $\alpha$ effectively introduces an angular deficit $2\pi(1 - \alpha)$ in the Minkowski geometry. 

The motion of the particle is governed by the Hamiltonian in cylindrical coordinates
\begin{equation}
H=-\frac{\hbar^2}{2m}\left[\frac{\partial^2}{\partial z^2}+\frac{1}{\rho}\frac{\partial}{\partial \rho}\left(\rho\frac{\partial}{\partial\rho}\right)+\frac{1}{\alpha^2\rho^2}
\frac{\partial^2}{\partial \theta^2}\right]. \label{hamiltonian}
\end{equation}

Since we are effectively excluding a cylindrical portion of the space accessible to the particle we must guarantee that the Hamiltonian is self-adjoint in the region of movement for, even if $H^{\dagger}=H$, their domains might be different. The von Neumann-Krein method \cite{reed} is used to find the self-adjoint extensions where needed. 

An operator $H$ with domain $D(H)$ is self-adjoint if $D(H^{\dagger})=D(H)$ and $H^{\dagger}=H$. The deficiency subspaces $N_{\pm}$  are  \cite{french}
\begin{eqnarray*}
N_{+}=\{ \psi \in D(H^{\dagger}), H^{\dagger}\psi=z_{+}\psi, Im\, z_{+}>0\},\\
N_{-}=\{ \psi \in D(H^{\dagger}), H^{\dagger}\psi=z_{-}\psi, Im\, z_{-}<0\},\\  
\end{eqnarray*}
with dimensions $n_{+}$ and $n_{-}$, respectively, which are called deficiency indices of $H$.  A necessary and sufficient condition  for $H$ to be self-adjoint is that $n_{+}=n_{-}=0$. On the other hand, if $n_{+}=n_{-}\geq 1$ then $H$ has an infinite number of self-adjoint extensions parametrized by a unitary $n\times n$ matrix, where $n=n_{+}=n_{-}$.

It is obvious that $(n_+,n_-)=(0,0)$ for $-\frac{\partial ^2}{\partial z^2}$ on $L^2(\mathbb R)$, the Hilbert space of square integrable functions on $\mathbb R$. Therefore $-\frac{\partial ^2}{\partial z^2}$ is essentially self-adjoint. This being considered, we have also that  translational invariance along the string axis ($z$-direction) reduces the problem to effectively two spacial dimensions.

Since the  typical size of a cosmic string radius is of the order of the GUT
length scale ($10^{-30}$cm) we can make the radius of the shielding cylinder $a\rightarrow 0$ without loss of generality. Effectively, we have ${\mathbb R}^3$ minus a line endowed with the space part of the metric (\ref{metric}).  We can then generalize the results of Kowalski {\it at al.} \cite{kowalski}, which studied the quantum dynamics of a free particle on a plane minus a point, to include the angular deficit which characterizes the cosmic string spacetime.

Following reference \cite{kowalski} we require that the wavefunctions must obey the periodicity condition
\begin{equation}
\psi(\theta+2\pi)=e^{i\lambda2\pi}\psi(\theta), \label{period}
\end{equation}
where $\lambda\in[0,1)$. Under this condition, $(n_+,n_-)=(0,0)$ for $-\frac{\partial ^2}{\partial \theta^2}$ on $L^2(S^1,d\theta)_{\lambda/\alpha}$, the Hilbert space of square integrable functions on the circle satisfying (\ref{period}) and with scalar product defined by
\begin{equation}
(\varphi,\chi)=\frac{1}{2\pi}\int_{o}^{2\pi}d\theta\varphi^*(\theta)\chi(\theta).
\end{equation}
The spectrum of $-\frac{\partial ^2}{\partial \theta^2}$ is given by $(\lambda+l)^2$, where $l$ is an integer.

With the above results in mind we assume the eigenfunctions of (\ref{hamiltonian}) to be of the form
\begin{equation}
\Psi(z,\rho,\theta)=e^{ikz+i(\lambda+l)\theta}\psi(\rho).
\end{equation}
This takes us to the effective Hamiltonian
\begin{equation}
H_{\frac{\lambda}{\alpha},\frac{l}{\alpha},k}=-\frac{\hbar^2}{2m}\left[\frac{1}{\rho}\frac{\partial}{\partial \rho}\left(\rho\frac{\partial}{\partial\rho}\right)-\frac{(\lambda+l)^2}{\alpha^2\rho^2}-k^2\right]. \label{effhamilt}
\end{equation}

Now, in order to examine the  deficiency subspaces $N_{\pm}$, we need to find the solution to
\begin{equation}
H^{\dagger}_{\frac{\lambda}{\alpha},\frac{l}{\alpha},k}\Phi_{\pm}=\pm i \kappa \Phi_{\pm}.
\end{equation}
The solution to this equation may be presented in terms of the modified Bessel functions $I_{\nu}$ and $K_{\nu}$, as follows
\begin{equation}
\Phi_{\pm}(\rho)=C_{1} I_{\frac{\lambda+l}{\alpha}}\left( \frac{\rho}{\hbar}\sqrt{\mp i2m\kappa}\right) + C_{2} K_{\frac{\lambda+l}{\alpha}}\left( \frac{\rho}{\hbar}\sqrt{\mp i2m\kappa}\right), \label{eigenfunction}
\end{equation} 
which belong to $L^2(\mathbb R^{+},\rho d\rho)$, the Hilbert space of square integrable functions on the half-axis, with measure $\rho d\rho$, only if $\frac{\lambda+l}{\alpha}\in(-1,1)$. Out of this range, the deficiency indices $(n_+,n_-)=(0,0)$, implying that $H_{\frac{\lambda}{\alpha},\frac{l}{\alpha},k}$ is essentially self-adjoint.

Requiring time-reversal symmetry on (\ref{period}), we have that \cite{kowalski}
\begin{equation}
\hat{T}\psi(\theta+2\pi)=e^{-i\lambda2\pi}\hat{T}\psi(\theta) \label{reversal}.
\end{equation}
It follows that equations (\ref{period}) and (\ref{reversal}) are compatible only if $\lambda=0$ or $\lambda=\frac{1}{2}$, which correspond to periodic or antiperiodic functions, respectively. The $\lambda=0$ case implies $l=0$, since $l$ is integer, $\frac{\lambda+l}{\alpha}\in(-1,1)$ and $\alpha<1$.  When $\lambda=l=0$ the effective angular momentum $\frac{\lambda+l}{\alpha}$ is zero and therefore the eigenstates (\ref{eigenfunction}) do not depend on $\alpha$, which does not characterize the Aharonov-Bohm effect. Therefore, the most interesting case for us is when $\lambda=\frac{1}{2}$, which allows  $l=0,-1$. 

For $l=0$, $\lambda=\frac{1}{2}$, we have a family of self-adjoint extensions with energies given by \cite{kowalski}
\begin{equation}
E_{\frac{1}{2\alpha},0,k}=-\kappa\left[\frac{\cos \left(\frac{\eta}{2}+\frac{1}{2\alpha}\frac{\pi}{4}\right)}{\cos \left(\frac{\eta}{2}-\frac{1}{2\alpha}\frac{\pi}{4}\right)}\right]^{2\alpha} + \frac{\hbar^2}{2m}k^2,
\end{equation}
where $\eta$ is the extension parameter and the second term on the right hand side of this equation comes from the free movement along the $z$ axis. The dependence of the energy on $\alpha$ characterizes the bound state Aharonov-Bohm effect. What is somewhat surprising in this expression is that even though the angular momentum $l=0$  the energy still depends on the parameter $\alpha$.  This is not seen in the other cases previously studied \cite{valdir,kaluza}.

\acknowledgments{This work was partially supported by PRONEX/FAPESQ-PB, CNPq and
CAPES (PROCAD)}.

\end{document}